\RequirePackage[bookmarksnumbered,unicode,hyperfootnotes=false]{hyperref}
\documentclass[sigconf,nonacm]{acmart}

\usepackage{booktabs}

\usepackage{enumitem}
\usepackage{pgfplots}
\pgfplotsset{compat=1.18}
\definecolor{hcorapblue}{HTML}{2F6690}
\definecolor{hcorapamber}{HTML}{A65F00}
\hypersetup{hidelinks}
\setcopyright{none}

\DeclareMathOperator{\CONT}{CONT}
\DeclareMathOperator{\OT}{OT}
\DeclareMathOperator{\SIM}{SIM}
\DeclareMathOperator{\STAB}{STAB}

\renewcommand\footnotetextcopyrightpermission[1]{}

\begin{document}

\title[Continuity-First Home-Care Resource Allocation]{Continuity-First
Lexicographic Optimization for Home-Care Resource Allocation}

\author{Tuyen Van Kieu}
\affiliation{%
  \institution{VNU University of Engineering and Technology}
  \city{Hanoi}
  \country{Vietnam}}
\email{tuyenkv@vnu.edu.vn}

\author{Khanh Ngoc Do}
\affiliation{%
  \institution{VNU University of Engineering and Technology}
  \city{Hanoi}
  \country{Vietnam}}
\email{23020615@vnu.edu.vn}

\author{Khanh Van To}
\authornote{Corresponding author.}
\affiliation{%
  \institution{VNU University of Engineering and Technology}
  \city{Hanoi}
  \country{Vietnam}}
\email{khanhtv@vnu.edu.vn}
\renewcommand{\shortauthors}{Kieu et al.}

\begin{abstract}
Home-care allocation must balance continuity of care, caregiver overtime, and caregiver-service compatibility. The published formulation of the Home-Care Optimal Resource Allocation Problem (HCORAP) uses a weighted policy (Weighted) to combine these outcomes, allowing compatibility gains to offset poorer continuity or additional overtime. We introduce LEX-COS, a lexicographic objective for HCORAP that first minimizes continuity violations, then overtime, and finally maximizes compatibility. We give an exact MaxSAT implementation and construct HCORAP-LC, a reproducible 48-instance synthetic evaluation suite with limited caregiver capacity. Compared with Weighted, LEX-COS reduces both continuity violations and overtime in 42 of 48 instances, without increasing either measure in any instance, while yielding lower compatibility. In 36 instances, LEX-COS achieves better continuity and lower overtime than every Weighted-optimal allocation. On the separate 48-instance Original benchmark, Gurobi and CPLEX, two leading commercial MIP solvers, resolve every instance under both policies and are substantially faster than EvalMaxSAT\@. EvalMaxSAT also resolves the full benchmark under LEX-COS\@. Within EvalMaxSAT, replacing sorting networks with Totalizer improves runtime under Weighted, with no clear gain under LEX-COS\@. These results characterize the effects of giving continuity and overtime explicit priority in home-care allocation.
\end{abstract}


\maketitle

\section{Introduction}
\label{sec:intro}

Home-care allocation decides which caregiver performs each requested service and when that service is delivered. Every allocation must respect caregiver qualifications, availability, and workload limits. Providers evaluate feasible allocations according to three outcomes. First, continuity of care favors assigning as few caregivers as possible to repeated visits for the same patient. Second, workload above each caregiver's regular allowance should be limited. Third, each service should be assigned to a compatible caregiver.

The relative importance of these outcomes determines which allocation is preferred. A weighted policy combines them in one score and allows them to compensate for one another. For example, the policy may accept an additional caregiver for one patient if the resulting assignments gain enough compatibility points. A strict-priority policy instead prevents a later criterion from degrading an earlier one. Throughout this paper, the term \emph{optimization policy} refers to the rule used to rank feasible allocations.

Unceta et al.~\cite{UncetaEtAl2024} formulate the Home-Care Optimal Resource Allocation Problem (HCORAP) as weighted partial MaxSAT\@. Their formulation provides an exact Boolean model for service coverage, caregiver eligibility, availability, workload, and care quality. Its published weights allow compatibility gains to compensate for poorer continuity or additional overtime. We refer to this published policy as Weighted. A provider that assigns first priority to continuity needs a different allocation rule: achieve the best feasible continuity, use the least overtime compatible with it, and then obtain the best caregiver-service match. We call this continuity-first objective LEX-COS, following the order $\CONT \to \OT \to \SIM$. It retains flexibility when perfect continuity is infeasible and makes the role of each care criterion explicit.

The evaluation answers two questions:

\begin{description}[leftmargin=2em,labelwidth=3em,itemsep=2pt,
  font=\normalfont\bfseries]
  \item[RQ1] How does a strict continuity-first policy change continuity, overtime, and compatibility relative to the Weighted policy, and do these changes persist across alternative optimal solutions, priority orders, weights, and capacity settings?
  \item[RQ2] How does replacing the sorting-network encoding with Totalizer affect exact MaxSAT performance when the remaining formulation is unchanged?
\end{description}

The paper makes three contributions:
\begin{enumerate}[leftmargin=1.6em,itemsep=2pt,topsep=3pt]
  \item We introduce LEX-COS, a new lexicographic objective for HCORAP, and implement it in MaxSAT\@.  The $\CONT \to \OT \to \SIM$ order protects the best feasible continuity, minimizes the overtime needed to preserve it, and maximizes compatibility within those priorities.
  \item We construct HCORAP-LC, a reproducible 48-instance evaluation suite for studying care priorities under limited capacity. Paired solutions, an analysis of all Weighted-optimal allocations, continuity budgets, and controlled weight and capacity sweeps quantify which differences are unavoidable and how the objectives trade off.
  \item We provide a controlled EvalMaxSAT comparison of sorting networks and Totalizers under both optimization policies, followed by a comparison with Gurobi and CPLEX on the same HCORAP instances.  This evaluation establishes exact reference results under both policies and measures the effects of encoding and solver choice.
\end{enumerate}

\section{Related Work}
\label{sec:related}

Home-care optimization includes assignment, scheduling, and routing decisions~\cite{FikarHirsch2017}. Integrated mixed-integer programming (MIP) models can assign caregivers, schedule visits, and construct routes while controlling workload and continuity~\cite{CappaneraScutella2015}. Other exact models address synchronized services~\cite{QiuEtAl2022} or flexible service times~\cite{MosqueraEtAl2019}. HCORAP assigns qualified caregivers and feasible time slots to requested services over a weekly planning horizon, but it does not construct travel routes~\cite{UncetaEtAl2024}.

Home-care models use priorities to express different planning goals. Lanzarone and Matta~\cite{LanzaroneMatta2014} preserve existing nurse-patient assignments and minimize the largest nurse overtime, then the next largest, in lexicographic order. Mosquera et al.~\cite{MosqueraEtAl2019} study flexible task durations and the prioritization of care tasks. Malagodi et al.~\cite{MalagodiEtAl2021} distinguish strict and soft preference matching in a routing model. LEX-COS addresses the joint choice of continuity, total overtime, and compatibility in HCORAP\@. It first minimizes the total number of additional caregivers across repeated-service groups, allowing the best feasible continuity even when a single caregiver cannot cover every visit. It then reduces overtime and improves compatibility without giving up an earlier optimum. This ordering supports a direct comparison of continuity, overtime, and compatibility with the published Weighted objective.

In multi-objective Boolean optimization, established methods optimize objectives in a fixed priority order~\cite{MarquesSilvaEtAl2011}. Recent work extends MaxSAT to multiple objectives~\cite{JabsEtAl2024}. MaxSAT has also been applied to staff scheduling~\cite{DemirovicEtAl2019}. Building on these methods, we give the three HCORAP criteria an explicit order and examine how that order changes the selected allocation.

Commercial MIP solvers have also been used in home-care research. For example, de Aguiar et al.~\cite{DeAguiarEtAl2023} compare Gurobi and CPLEX for routing and scheduling synchronized caregiver teams. The HCORAP study of Unceta et al.~\cite{UncetaEtAl2024} evaluates MaxSAT solvers and selects EvalMaxSAT for its experiments. We extend that comparison to EvalMaxSAT, Gurobi, and CPLEX on a common HCORAP benchmark under both Weighted and LEX-COS\@. This comparison evaluates both care outcomes and the computational effort required to obtain exact allocations.

The original HCORAP formulation represents counting constraints with cardinality networks~\cite{AsinEtAl2011}. A Totalizer is a tree-structured alternative for representing the same constraints~\cite{BailleuxBoufkhad2003}. This representation choice can substantially change MaxSAT performance~\cite{MorgadoEtAl2014}. We examine this choice under both the published Weighted objective and the proposed priority order.

\section{Problem Formulation}
\label{sec:formulation}

\subsection{Allocation model}
\label{sec:model}

Let $\mathcal{A}$ be the set of caregivers, $\mathcal{S}$ the set of requested services, $\mathcal{H}$ the set of time slots, and $\mathcal{U}$ the set of patients. The set $\mathcal{Q}$ contains the continuity groups. Each continuity group collects repeated services of the same type for one patient, such as all wound-care visits requested by that patient during the planning horizon. The set $\mathcal{S}_u$ contains patient $u$'s services, and $\mathcal{S}_q$ contains the services in continuity group $q$.

The set $E$ contains every feasible assignment triple $(a,s,h)$. Such a triple indicates that caregiver $a$ is qualified and available to perform service $s$ in time slot $h$. The binary variable $x_{ash}$ equals one when the allocation selects this triple. The derived variable $y_{as}$ indicates whether caregiver $a$ performs service $s$:
\begin{equation*}
  y_{as}\leftrightarrow\bigvee_{h:(a,s,h)\in E}x_{ash}.
\end{equation*}
For each caregiver $a$, $R_a$ is the regular workload allowance, and $O_a$ is the maximum additional workload permitted as overtime.

Every valid allocation must satisfy the following feasibility constraints:
\begin{align}
  \textstyle\sum_{(a,h):(a,s,h)\in E} x_{ash} &= 1
    & \forall s \in \mathcal{S}, \label{eq:cover}\\
  \textstyle\sum_{s:(a,s,h)\in E} x_{ash} &\leq 1
    & \forall a,h, \label{eq:caregiver-slot}\\
  \textstyle\sum_a\sum_{s\in\mathcal{S}_u:(a,s,h)\in E} x_{ash}
    &\leq 1
    & \forall u,h, \label{eq:patient-slot}\\
  L_a = \textstyle\sum_s y_{as} &\leq R_a + O_a
    & \forall a. \label{eq:workload}
\end{align}

Equation~\eqref{eq:cover} covers every service exactly once. Equations~\eqref{eq:caregiver-slot} and~\eqref{eq:patient-slot} prevent a caregiver from delivering, or a patient from receiving, two simultaneous services. Equation~\eqref{eq:workload} defines the workload $L_a$ and keeps it within the caregiver's total allowance. The benchmark records workload in service units, so overtime is reported in the same unit rather than in clock hours.

\subsection{Objective functions}
\label{sec:measures}

The indicator $v_{aq}$ equals one when caregiver $a$ performs at least one service in continuity group $q$, and $n_q$ counts the distinct caregivers used in that group:
\begin{equation*}
  v_{aq}\leftrightarrow\bigvee_{s\in\mathcal{S}_q}y_{as},
  \qquad n_q=\sum_a v_{aq}.
\end{equation*}
The parameter \mbox{$r_{as}\in\{0,\ldots,4\}$} measures the compatibility between caregiver $a$ and service $s$. The original HCORAP data call this quantity similarity, so we use the symbol $\SIM$ for the total compatibility score. The three allocation measures are defined below. We minimize $\CONT$ and $\OT$ and maximize $\SIM$:
\begin{align}
  \CONT &= \textstyle\sum_{q\in\mathcal{Q}}\max(0,\,n_q-1),
    \label{eq:cont}\\
  \OT   &= \textstyle\sum_{a\in\mathcal{A}}\max(0,\,L_a-R_a),
    \label{eq:ot}\\
  \SIM  &= \textstyle\sum_{a,s} r_{as}\,y_{as}.
    \label{eq:sim}
\end{align}

The value $\CONT$ counts additional caregivers beyond the first across continuity groups; each additional caregiver contributes one continuity violation. Thus, a lower value means better continuity. The value $\OT$ counts workload above regular allowances, and $\SIM$ sums the compatibility scores of the selected caregiver-service pairs. Under full service coverage, the original stability reward satisfies
\begin{equation*}
  \STAB+\CONT=\sum_{q\in\mathcal Q}(|\mathcal S_q|-1).
\end{equation*}
The sum is fixed, so minimizing $\CONT$ maximizes $\STAB$.

\subsection{Weighted and lexicographic objectives}
\label{sec:lex}

Up to an instance constant, the Weighted policy maximizes
\begin{equation}
  \SIM-w_c\,\CONT-w_o\,p\,\OT.
  \label{eq:weighted}
\end{equation}
The coefficients $w_c$ and $w_o$ weight continuity and overtime, respectively, and $p$ is the magnitude of the overtime penalty stored in the instance. We use the published setting $(w_c,w_o)=(1,1)$.

LEX-COS instead uses the strict order
\begin{equation}
  \min\,\CONT
    \;\longrightarrow\;
    \min\,\OT
    \;\longrightarrow\;
    \max\,\SIM.
  \label{eq:lex-cos}
\end{equation}

Each stage preserves the optimum already attained. The policy therefore achieves the best feasible continuity with the least overtime, then selects the most compatible allocation among those meeting both priorities. Unlike Weighted, LEX-COS does not require coefficients that convert measures with different units into a common score. The criterion order alone determines which trade-offs are admissible.

For sensitivity analysis, we also solve the overtime-first order $\min\OT\to\min\CONT\to\max\SIM$, denoted LEX-OT\@. Comparing the two orders measures the overtime needed to protect continuity when the separate continuity and overtime optima conflict.

\section{MaxSAT Encoding}
\label{sec:encoding}

The MaxSAT formulation represents the feasibility constraints in Section~\ref{sec:formulation} as mandatory clauses. Weighted assigns soft-clause costs that combine the three quality measures. LEX-COS solves three successive MaxSAT stages while retaining the best value found at each earlier stage. For example, after the first stage obtains the continuity optimum $C^*$, the second stage minimizes overtime only among allocations with $\CONT=C^*$. The final stage likewise maximizes compatibility among allocations that attain both earlier optima.

The formulation counts caregivers within continuity groups and workload per caregiver. These counts serve two purposes: they enforce workload limits and determine continuity violations and overtime. A \emph{cardinality encoding} represents these counts with clauses that a MaxSAT solver can process. The original \emph{sorting network} (SN)~\cite{AsinEtAl2011} uses a comparator circuit to produce the counts required by the formulation.

We select a \emph{Totalizer tree} (TOT)~\cite{BailleuxBoufkhad2003} as an alternative because it provides the same counts for both the constraints and objectives, allowing us to retain the remaining formulation. Totalizer splits the inputs into groups and combines their counts along a tree. This changes the auxiliary variables and clauses through which the solver reasons about continuity and workload, which can affect solving time and memory use. The comparison therefore examines whether changing the representation of these recurring counts improves MaxSAT performance. Testing both Weighted and LEX-COS also shows whether an encoding's benefit persists when the same care measures are optimized with different priorities.

\section{Experiments}
\label{sec:experiments}

\subsection{Test instances}

We use two 48-instance suites for different purposes. Each suite combines two patient counts (30 and 40), four caregiver counts (10, 15, 20, and 25), and two service counts per patient (4 and 5). The resulting 16 size classes contain three seeds each.

The \emph{Original} suite follows Unceta et al.~\cite{UncetaEtAl2024} and supports direct comparison with the published MaxSAT experiments. The \emph{HCORAP-LC} suite retains the same size classes, uses three new random seeds, and increases the load relative to Original to study policy trade-offs under limited caregiver capacity. We calibrate the load ratio to approximately 0.85, with
\mbox{$\rho=|\mathcal S|/\sum_a(R_a+O_a)$}.  This
ratio compares the number of requested services with the total regular and overtime capacity. Regular capacity accounts for approximately 85\% of this total. The suite is generated using an adaptation of the published generator. For each patient count and seed, the caregiver-count and service-count variants share one common base population. Every generated instance includes a verified feasible allocation.

\subsection{Evaluation protocol}

We compare Weighted and LEX-COS on all 48 HCORAP-LC instances, using Gurobi to establish exact reference optima. We also solve LEX-OT on all instances and cross-check all three policies with CPLEX on a 16-instance subset covering every size class. Every run has a 300-second limit.

To examine all Weighted-optimal allocations, we fix the optimal Weighted score and find the smallest and largest $\CONT$ and $\OT$ values attainable at that score. Together, these bounds describe the range of each measure among all Weighted-optimal allocations. If the smallest attainable value exceeds its LEX-COS counterpart, every Weighted-optimal allocation is worse on that criterion. We also allow $\CONT$ to exceed its best value by a budget of $k\in\{0,1,2\}$, then minimize $\OT$ and maximize $\SIM$. The value $k$ is the permitted increase in continuity violations above the optimum. At $k=0$, the procedure recovers the LEX-COS quality vector; larger budgets show how relaxing continuity affects overtime and compatibility.

Two further matched sweeps test sensitivity. The weight sweep uses $w_c,w_o\in\{1,4,8\}$ while keeping the compatibility coefficient equal to one. The capacity sweep solves Weighted, LEX-COS, and LEX-OT for nine combinations of target load $\rho\in\{0.55,0.85,0.98\}$ and regular-capacity fraction in $\{0.70,0.85,1.00\}$. Each variant changes only regular and overtime allowances and retains a verified feasible allocation. The capacity variants are nested within 6 generated patient-seed families; we report descriptive matched comparisons within these families.

To compare encodings and solvers, we evaluate four configurations on each of the 48 Original instances: Weighted with SN, Weighted with Totalizer, LEX-COS with SN, and LEX-COS with Totalizer. We use EvalMaxSAT, the strongest MaxSAT solver tested in the original HCORAP study~\cite{UncetaEtAl2024,Avellaneda2024EvalMaxSAT}. Gurobi and CPLEX also solve both policies on every instance, using the same hardware configuration, thread count, and time limit. Every run has a 3,600-second budget, shared across all stages for LEX-COS\@. Paired encoding runs use the same solver configuration, and their order is randomized within each instance.

\subsection{Experimental setup}

All measured runs use one thread on a Google Cloud virtual machine with 8 vCPUs and 16 GB RAM\@. We call a run
\emph{proved} when the solver proves either an optimal solution or
infeasibility. A LEX-COS run is optimal only when all three stages finish and prove their respective optima.

PAR-2 is the average runtime after every unfinished run is assigned a penalty of twice the limit, $2T$. We also report the numbers of optimal, infeasible, and timed-out runs and the median runtime over proved runs. A paired runtime comparison includes only instances for which both encoding configurations produce a proved result. For each such instance, the percentage runtime reduction is $100(t_{\mathrm{SN}}-t_{\mathrm{TOT}})/t_{\mathrm{SN}}$, where $t_{\mathrm{SN}}$ and $t_{\mathrm{TOT}}$ are the respective EvalMaxSAT runtimes with the sorting-network and Totalizer encodings on the same instance under the same policy. We report the median reduction and its 95\% bootstrap confidence interval, resampling complete instance pairs.

An independent verification script checks every reported allocation and recomputes $\CONT$, $\OT$, and $\SIM$.

\section{Results}
\label{sec:results}

\subsection{RQ1: Comparison of optimization policies}
\label{sec:rq1}

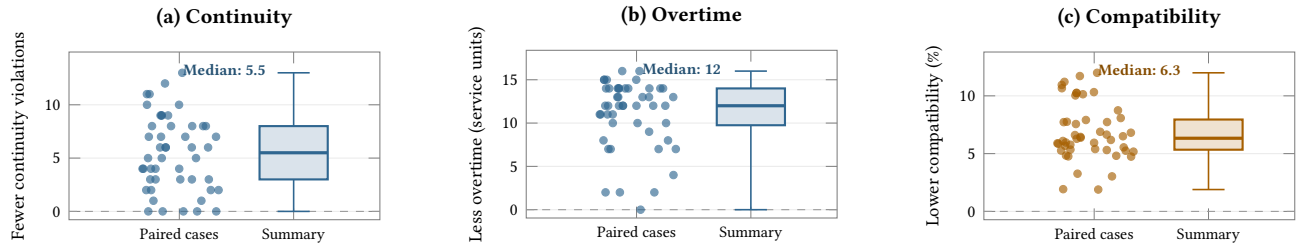
\begin{figure*}[t]
\centering
\begin{minipage}[t]{0.32\textwidth}\centering
\begin{tikzpicture}\begin{axis}[
width=\linewidth,height=3.8cm,xmin=0,xmax=1.55,
xtick={0.55,1.12},xticklabels={Paired cases,Summary},
ymin=-0.78000000,ymax=14.95000000,
title={(a) Continuity},ylabel={Fewer continuity violations},
title style={font=\small\bfseries},label style={font=\footnotesize},
tick label style={font=\scriptsize},ymajorgrids,grid style={black!8},
axis line style={black!50}]
\draw[dashed,black!40] (axis cs:0,0)--(axis cs:1.55,0);
\addplot[only marks,mark=*,mark size=1.4pt,hcorapblue,opacity=0.65] coordinates {(0.570226,-0.00000000) (0.395394,-0.00000000) (0.638850,-0.00000000) (0.466945,-0.00000000) (0.734097,-0.00000000) (0.420942,1.00000000) (0.660034,1.00000000) (0.409569,2.00000000) (0.745757,2.00000000) (0.706916,2.00000000) (0.385158,2.00000000) (0.612713,3.00000000) (0.431581,3.00000000) (0.554872,3.00000000) (0.697130,3.00000000) (0.405065,3.00000000) (0.431184,4.00000000) (0.370794,4.00000000) (0.366672,4.00000000) (0.549532,4.00000000) (0.408475,4.00000000) (0.394112,5.00000000) (0.633221,5.00000000) (0.462536,5.00000000) (0.482417,6.00000000) (0.481855,6.00000000) (0.693481,6.00000000) (0.454204,6.00000000) (0.612184,6.00000000) (0.733027,7.00000000) (0.579879,7.00000000) (0.458823,7.00000000) (0.398037,7.00000000) (0.614173,8.00000000) (0.413759,8.00000000) (0.508735,8.00000000) (0.682376,8.00000000) (0.668658,8.00000000) (0.456447,9.00000000) (0.493866,9.00000000) (0.461388,9.00000000) (0.462378,9.00000000) (0.548720,10.00000000) (0.388907,10.00000000) (0.388342,11.00000000) (0.401700,11.00000000) (0.478585,12.00000000) (0.564690,13.00000000)};
\draw[hcorapblue,semithick] (axis cs:1.12,-0.0)--(axis cs:1.12,13.0);
\draw[hcorapblue,semithick] (axis cs:1.04,-0.0)--(axis cs:1.20,-0.0);
\draw[hcorapblue,semithick] (axis cs:1.04,13.0)--(axis cs:1.20,13.0);
\filldraw[fill=hcorapblue!18,draw=hcorapblue,thick] (axis cs:0.95,3.0) rectangle (axis cs:1.29,8.0);
\draw[hcorapblue,very thick] (axis cs:0.95,5.5)--(axis cs:1.29,5.5);
\node[anchor=north,font=\scriptsize\bfseries,text=hcorapblue!80!black] at (rel axis cs:0.5,0.98) {Median: 5.5};
\end{axis}\end{tikzpicture}\end{minipage}
\hfill
\begin{minipage}[t]{0.32\textwidth}\centering
\begin{tikzpicture}\begin{axis}[
width=\linewidth,height=3.8cm,xmin=0,xmax=1.55,
xtick={0.55,1.12},xticklabels={Paired cases,Summary},
ymin=-0.96000000,ymax=18.40000000,
title={(b) Overtime},ylabel={Less overtime (service units)},
title style={font=\small\bfseries},label style={font=\footnotesize},
tick label style={font=\scriptsize},ymajorgrids,grid style={black!8},
axis line style={black!50}]
\draw[dashed,black!40] (axis cs:0,0)--(axis cs:1.55,0);
\addplot[only marks,mark=*,mark size=1.4pt,hcorapblue,opacity=0.65] coordinates {(0.570226,-0.00000000) (0.395394,2.00000000) (0.638850,2.00000000) (0.466945,2.00000000) (0.734097,4.00000000) (0.420942,7.00000000) (0.660034,7.00000000) (0.409569,7.00000000) (0.745757,7.00000000) (0.706916,8.00000000) (0.385158,8.00000000) (0.612713,9.00000000) (0.431581,10.00000000) (0.554872,10.00000000) (0.697130,10.00000000) (0.405065,11.00000000) (0.431184,11.00000000) (0.370794,11.00000000) (0.366672,11.00000000) (0.549532,12.00000000) (0.408475,12.00000000) (0.394112,12.00000000) (0.633221,12.00000000) (0.462536,12.00000000) (0.482417,12.00000000) (0.481855,12.00000000) (0.693481,12.00000000) (0.454204,13.00000000) (0.612184,13.00000000) (0.733027,13.00000000) (0.579879,13.00000000) (0.458823,13.00000000) (0.398037,14.00000000) (0.614173,14.00000000) (0.413759,14.00000000) (0.508735,14.00000000) (0.682376,14.00000000) (0.668658,14.00000000) (0.456447,14.00000000) (0.493866,14.00000000) (0.461388,14.00000000) (0.462378,14.00000000) (0.548720,15.00000000) (0.388907,15.00000000) (0.388342,15.00000000) (0.401700,15.00000000) (0.478585,16.00000000) (0.564690,16.00000000)};
\draw[hcorapblue,semithick] (axis cs:1.12,-0.0)--(axis cs:1.12,16.0);
\draw[hcorapblue,semithick] (axis cs:1.04,-0.0)--(axis cs:1.20,-0.0);
\draw[hcorapblue,semithick] (axis cs:1.04,16.0)--(axis cs:1.20,16.0);
\filldraw[fill=hcorapblue!18,draw=hcorapblue,thick] (axis cs:0.95,9.75) rectangle (axis cs:1.29,14.0);
\draw[hcorapblue,very thick] (axis cs:0.95,12.0)--(axis cs:1.29,12.0);
\node[anchor=north,font=\scriptsize\bfseries,text=hcorapblue!80!black] at (rel axis cs:0.5,0.98) {Median: 12};
\end{axis}\end{tikzpicture}\end{minipage}
\hfill
\begin{minipage}[t]{0.32\textwidth}\centering
\begin{tikzpicture}\begin{axis}[
width=\linewidth,height=3.8cm,xmin=0,xmax=1.55,
xtick={0.55,1.12},xticklabels={Paired cases,Summary},
ymin=-0.71959459,ymax=13.79222973,
title={(c) Compatibility},ylabel={Lower compatibility (\%)},
title style={font=\small\bfseries},label style={font=\footnotesize},
tick label style={font=\scriptsize},ymajorgrids,grid style={black!8},
axis line style={black!50}]
\draw[dashed,black!40] (axis cs:0,0)--(axis cs:1.55,0);
\addplot[only marks,mark=*,mark size=1.4pt,hcorapamber,opacity=0.65] coordinates {(0.570226,1.89125296) (0.395394,1.92307692) (0.638850,3.03030303) (0.466945,3.26340326) (0.734097,4.74040632) (0.420942,4.76190476) (0.660034,4.81283422) (0.409569,4.83333333) (0.745757,5.17241379) (0.706916,5.25291829) (0.385158,5.26315789) (0.612713,5.30973451) (0.431581,5.34759358) (0.554872,5.39629005) (0.697130,5.54272517) (0.405065,5.68181818) (0.431184,5.76923077) (0.370794,5.86011342) (0.366672,5.89198036) (0.549532,5.93869732) (0.408475,5.96330275) (0.394112,6.09480813) (0.633221,6.18421053) (0.462536,6.28272251) (0.482417,6.37813212) (0.481855,6.44599303) (0.693481,6.50224215) (0.454204,6.60592255) (0.612184,6.64335664) (0.733027,6.81003584) (0.579879,6.88888889) (0.458823,7.56062767) (0.398037,7.72471910) (0.614173,7.73109244) (0.413759,7.75623269) (0.508735,7.90861160) (0.682376,8.07799443) (0.668658,8.74316940) (0.456447,10.03401361) (0.493866,10.13745704) (0.461388,10.23765996) (0.462378,10.26666667) (0.548720,10.32171582) (0.388907,10.65719361) (0.388342,10.94594595) (0.401700,11.21157324) (0.478585,11.71171171) (0.564690,11.99324324)};
\draw[hcorapamber,semithick] (axis cs:1.12,1.8912529550827424)--(axis cs:1.12,11.993243243243244);
\draw[hcorapamber,semithick] (axis cs:1.04,1.8912529550827424)--(axis cs:1.20,1.8912529550827424);
\draw[hcorapamber,semithick] (axis cs:1.04,11.993243243243244)--(axis cs:1.20,11.993243243243244);
\filldraw[fill=hcorapamber!18,draw=hcorapamber,thick] (axis cs:0.95,5.338128815484359) rectangle (axis cs:1.29,7.9509573067151);
\draw[hcorapamber,very thick] (axis cs:0.95,6.330427315770015)--(axis cs:1.29,6.330427315770015);
\node[anchor=north,font=\scriptsize\bfseries,text=hcorapamber!80!black] at (rel axis cs:0.5,0.98) {Median: 6.3};
\end{axis}\end{tikzpicture}\end{minipage}
\caption{Paired changes from Weighted to LEX-COS are shown for 48 HCORAP-LC instances.
Dots show instances, and box plots summarize their distributions. Panels (a) and (b) show improvements; panel (c) shows compatibility reduction.}
\label{fig:policy-effect}
\Description{Three distributions show fewer continuity violations, less overtime,
and the percentage reduction in compatibility under LEX-COS.}
\end{figure*}

On the 48 HCORAP-LC instances, LEX-COS improves both continuity and overtime in 42 cases and worsens neither measure. It reduces continuity violations in
43 of the 48 instances and reduces overtime
in 47. The median reductions are
5.5 continuity violations and
12 overtime workload units.  Compatibility is lower
in all 48 pairs, with a median reduction of
6.3\% relative to each instance's Weighted
compatibility score. Figure~\ref{fig:policy-effect} shows the gains in continuity and overtime alongside the reduction in compatibility.

Examining all Weighted-optimal allocations separates the effect of the objective from the choice among equally scoring allocations. Relative to LEX-COS, the best continuity value among these allocations remains worse in
41/48 instances,
and the best overtime value remains worse in
43/48 instances.
Both values remain worse in
36/48 instances.

Reversing the first two priorities changes the reported quality vector in 3 of the 48 HCORAP-LC instances. In each of these cases, LEX-COS removes one continuity violation but uses one additional overtime unit. The two priority orders return the same quality values in the other 45 instances, where continuity and overtime can attain their separate best values together. The budget analysis quantifies the effects of relaxing continuity. Allowing one additional continuity violation reduces overtime in
3 instances, by
3 units in total, and increases the
compatibility score by a median of 8.

Increasing both penalty weights together gives continuity and overtime more influence relative to compatibility while preserving the trade-off between them. The settings $(1,1)$, $(4,4)$, and $(8,8)$ match the complete LEX-COS quality vector in
0, 8, and
25 returned solutions out of 48, respectively.
With weights $(8,1)$, the returned solutions match the LEX-COS continuity value in 32 instances, but none matches its complete quality vector. No tested weight setting reproduces the LEX-COS quality vector across all 48 instances.

Across the nine capacity settings, Weighted and LEX-COS differ on at least
40 and at most 48 instances
per setting. At the capacity setting that matches HCORAP-LC's load and regular-capacity fraction, the two priority orders differ on 3 instances. The repository reports the complete budget, weight, and capacity results.

\subsection{RQ2: Comparison of cardinality encodings}
\label{sec:rq2}

\begin{table}[tbp]
\caption{EvalMaxSAT performance by policy and encoding on 48 Original instances.}
\label{tab:policy-encoding}
\centering
\small
\setlength{\tabcolsep}{3pt}
\renewcommand{\arraystretch}{1.15}
\begin{tabular*}{\linewidth}{@{\extracolsep{\fill}}llccc@{}}
\toprule
 & & Result counts & \multicolumn{2}{c}{Runtime (s)}\\
\cmidrule(lr){3-3}\cmidrule(l){4-5}
Policy & Encoding & Opt / Inf / TO & PAR-2 & Median\\
\midrule
Weighted & SN & 40 / 6 / 2 & 352.9 & 30.1\\
Weighted & TOT & 40 / 6 / 2 & \textbf{333.0} & \textbf{26.9}\\
\addlinespace
LEX-COS & SN & \textbf{42 / 6 / 0} & \textbf{95.8} & 97.0\\
LEX-COS & TOT & \textbf{42 / 6 / 0} & 97.3 & \textbf{93.9}\\
\bottomrule
\end{tabular*}
\par\smallskip\begin{minipage}{\linewidth}\footnotesize
Opt/Inf/TO: optimal, infeasible, timed-out. Median over proved runs; PAR-2 over all 48 runs (penalty $2T$).
\end{minipage}
\end{table}

Both encodings produce the same numbers of optimal, infeasible, and timed-out runs within each policy, but their runtime difference depends on the policy. Table~\ref{tab:policy-encoding} reports solution status and runtime. Under Weighted, among the 46 instances for which EvalMaxSAT proves a result with both encodings, Totalizer is faster on
35.  The median per-instance runtime reduction is
14.0\%, with a 95\% confidence interval
of [8.7\%, 19.0\%].

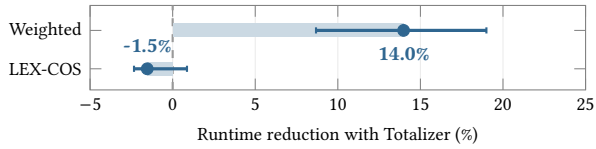
\begin{figure}[tbp]\centering
\begin{tikzpicture}\begin{axis}[
width=0.96\linewidth,height=2.7cm,
xmin=-5,xmax=25,ymin=0.45,ymax=2.65,
xtick={-5,0,5,10,15,20,25},
ytick={1,2},yticklabels={LEX-COS,Weighted},
xlabel={Runtime reduction with Totalizer (\%)},
label style={font=\footnotesize},tick label style={font=\footnotesize},
xmajorgrids,grid style={black!8},axis line style={black!50}]
\draw[dashed,black!40,thick] (axis cs:0,0.45)--(axis cs:0,2.65);
\fill[hcorapblue!25] (axis cs:0.0,1.82) rectangle (axis cs:13.982434488212085,2.18);
\draw[hcorapblue,very thick] (axis cs:8.679288129206046,2)--(axis cs:18.99618222064876,2);
\draw[hcorapblue,thick] (axis cs:8.679288129206046,1.92)--(axis cs:8.679288129206046,2.08);
\draw[hcorapblue,thick] (axis cs:18.99618222064876,1.92)--(axis cs:18.99618222064876,2.08);
\addplot[mark=*,mark size=2.2pt,hcorapblue] coordinates {(13.982434488212085,2)};
\node[anchor=north,yshift=-3pt,font=\small\bfseries,hcorapblue] at (axis cs:13.982434488212085,2) {14.0\%};
\fill[hcorapblue!25] (axis cs:-1.5404259352007243,0.8200000000000001) rectangle (axis cs:0.0,1.18);
\draw[hcorapblue,very thick] (axis cs:-2.3401068045648667,1)--(axis cs:0.8671879287716313,1);
\draw[hcorapblue,thick] (axis cs:-2.3401068045648667,0.92)--(axis cs:-2.3401068045648667,1.08);
\draw[hcorapblue,thick] (axis cs:0.8671879287716313,0.92)--(axis cs:0.8671879287716313,1.08);
\addplot[mark=*,mark size=2.2pt,hcorapblue] coordinates {(-1.5404259352007243,1)};
\node[anchor=south,yshift=3pt,font=\small\bfseries,hcorapblue] at (axis cs:-1.5404259352007243,1) {-1.5\%};
\end{axis}\end{tikzpicture}
\caption{Bars show median paired runtime reductions from SN to Totalizer,
with 95\% bootstrap confidence intervals.}
\label{fig:encoding-effect}
\Description{Horizontal bars show the median percentage runtime reduction
under each policy, with confidence intervals and a zero baseline.}
\end{figure}

Under LEX-COS, EvalMaxSAT resolves all 48 instances with both encodings. Totalizer is faster on
20 of the 48
pairs. Totalizer is slower by a median of
1.5\%; the 95\% confidence interval for its
runtime reduction is [-2.3\%,
0.9\%] and includes zero.
Figure~\ref{fig:encoding-effect} summarizes this policy-dependent effect.

Totalizer reduces PAR-2 by 5.6\% under Weighted but increases it by 1.5\% under LEX-COS\@. The encoding effect also varies across size classes; the repository reports the stratified results. Under LEX-COS, Totalizer uses 3.7 times the median peak memory of SN\@. The repository also reports formula sizes and memory use.

\subsection{Comparison of exact solvers}
\label{sec:validation}

\begin{table}[tbp]
\caption{Exact-solver performance on 48 Original instances per policy.}
\label{tab:exact-solvers}
\centering
\small
\setlength{\tabcolsep}{3pt}
\renewcommand{\arraystretch}{1.12}
\begin{tabular*}{\columnwidth}{@{\extracolsep{\fill}}llccc@{}}
\toprule
 & & Result counts & \multicolumn{2}{c}{Runtime (s)}\\
\cmidrule(lr){3-3}\cmidrule(l){4-5}
Solver & Policy & Opt / Inf / TO & PAR-2 & Median\\
\midrule
EvalMaxSAT & Weighted & 40 / 6 / 2 & 333.0 & 26.9\\
EvalMaxSAT & LEX-COS & \textbf{42 / 6 / 0} & 97.3 & 93.9\\
\addlinespace
Gurobi & Weighted & \textbf{42 / 6 / 0} & \textbf{0.09} & \textbf{0.08}\\
Gurobi & LEX-COS & \textbf{42 / 6 / 0} & \textbf{0.23} & \textbf{0.22}\\
\addlinespace
CPLEX & Weighted & \textbf{42 / 6 / 0} & 0.17 & 0.13\\
CPLEX & LEX-COS & \textbf{42 / 6 / 0} & 0.46 & 0.45\\
\bottomrule
\end{tabular*}
\par\smallskip\begin{minipage}{\linewidth}\footnotesize
Opt/Inf/TO: optimal, infeasible, timed-out. EvalMaxSAT uses Totalizer. Median over proved runs; PAR-2 over all 48 runs (penalty $2T$).
\end{minipage}
\end{table}

Table~\ref{tab:exact-solvers} compares EvalMaxSAT with the two MIP solvers on the 48 Original instances under both policies, using Totalizer as the fixed MaxSAT representation. Gurobi and CPLEX resolve all 96 instance-policy combinations, comprising 84 optimal and 12 infeasible cases. They agree on every status and on all quality values in the optimal cases; their results also agree with every proved EvalMaxSAT result. Both solvers prove optimality for the two Weighted cases that time out in EvalMaxSAT\@. Their subsecond median runtimes are substantially lower than those of EvalMaxSAT on this benchmark.

\section{Discussion}
\label{sec:discussion}

LEX-COS prevents later criteria from compensating for worse continuity or overtime. Its gains persist across alternative Weighted-optimal allocations, showing the effect of the objective itself.

Continuity and overtime attain their separate optima together in most capacity settings. In the three observed conflicts, one fewer continuity violation requires one additional overtime unit. The budget analysis shows how relaxing continuity can reduce overtime and recover compatibility. Stronger penalties bring Weighted closer to LEX-COS without reliably matching its complete quality vector.

Within EvalMaxSAT, Totalizer improves runtime under Weighted but shows no clear gain under LEX-COS, for which SN also uses less memory. These results show that an encoding choice that benefits one optimization policy need not benefit another. On the Original benchmark, Gurobi is fastest, followed by CPLEX; both have subsecond median runtimes. These findings apply to the tested instance sizes, solver settings, and hardware.

The synthetic instances model weekly allocations with fixed time slots and known availability but no travel. Workload is measured in service units, and compatibility represents assignment preferences. Future work with operational data can address variable visit durations, routing, uncertain demand, and care outcomes.

\section{Conclusion}
\label{sec:conclusion}
LEX-COS makes the care priorities explicit by optimizing continuity, overtime, and compatibility in sequence. On HCORAP-LC, it improves continuity and overtime in 42/48 instances, while lower compatibility exposes the cost of those priorities. The Weighted-optimal allocations, continuity budgets, and sensitivity studies identify which differences follow from the objective and when the first two priorities conflict. LEX-COS thus provides an exact allocation policy for providers that prioritize continuity, then overtime, and finally compatibility.

Gurobi and CPLEX resolve every tested Original instance and are faster than EvalMaxSAT, which resolves the full benchmark under LEX-COS\@. Encoding performance also depends on the policy, linking care priorities to both the selected allocation and computational behavior. Evaluation with operational data and travel-aware schedules can test whether these findings extend to broader home-care planning settings.

\bibliographystyle{ACM-Reference-Format}
\bibliography{references}

\end{document}